\documentclass[graybox]{svmult}

\usepackage{mathptmx}       
\usepackage{helvet}         
\usepackage{courier}        
\usepackage{type1cm}        
\usepackage{makeidx}         
\usepackage{graphicx}        
\usepackage{multicol}        
\usepackage[bottom]{footmisc}

\usepackage{amsmath,amssymb}

\newcommand{\dx}{d^4x}

\newcommand{\nmu}{\nabla_{\mu}}
\newcommand{\nnu}{\nabla_{\nu}}

\newcommand{\FF}{\mathcal{F}}

\newcommand{\be}{\begin{equation}}
\newcommand{\ee}{\end{equation}}
\newcommand{\bea}{\begin{eqnarray}}
\newcommand{\eea}{\end{eqnarray}}
\newcommand{\pd}{\partial}

\begin{document}

\title*{On Nonlocal Modified Gravity \\ and its Cosmological Solutions}
\author{Ivan Dimitrijevic, Branko Dragovich, Jelena Stankovic, Alexey S. Koshelev \\ and Zoran Rakic}
\authorrunning{I. Dimitrijevic et al.}
\institute{I. Dimitrijevic \at Faculty of Mathematics, University of Belgrade,
Studentski trg 16,  Belgrade, Serbia \\ \email{ivand@matf.bg.ac.rs}\and
B. Dragovich \at Institute of Physics,\, University of Belgrade; Mathematical Institute SANU,
Belgrade,\, Serbia \\ \email{dragovich@ipb.ac.rs} \and
J. Stankovic \at Teacher Education Faculty, University of Belgrade,  Kraljice Natalije 43, Belgrade, Serbia \\ \email{jelenagg@gmail.com}
\and A. S. Koshelev \at Departamento  de F\'isica and Centro  de  Matem\'atica  e
Aplica\c c\~oes,  Universidade  da  Beira  Interior,  6200  Covilh\~a,
Portugal; Theoretische Natuurkunde, Vrije Universiteit Brussel, and The
International Solvay Institutes, Pleinlaan 2, B-1050 Brussels, Belgium
 \\ \email{alexey@ubi.pt}
\and Z. Rakic \at Faculty of Mathematics, University of Belgrade,
Studentski trg 16,  Belgrade, Serbia \\ \email{zrakic@matf.bg.ac.rs}
}

%
%
\maketitle

\abstract{During hundred years of General Relativity (GR), many significant gravitational phenomena have been predicted and discovered. General Relativity is still the best theory of  gravity. Nevertheless, some (quantum) theoretical and (astrophysical and cosmological) phenomenological difficulties of modern gravity have been motivation to search more general theory  of gravity than GR. As a result, many modifications of GR have been considered. One of promising  recent investigations is Nonlocal Modified Gravity. In this article we present a brief review of some nonlocal gravity models with their cosmological solutions, in which nonlocality is expressed by an analytic function of the d'Alembert-Beltrami
operator $\Box$. Some new results are also presented. }


\section{Introduction}
\label{sec:1}
General relativity (GR) was formulated one hundred years ago and is also known as Einstein theory of gravity.
GR is regarded as one of the most profound and beautiful physical theories with great phenomenological  achievements and nice theoretical properties.
It has been tested and quite well confirmed in the Solar system, and it has been also used as a theoretical laboratory
for gravitational investigations at other spacetime scales. GR has important astrophysical implications predicting
existence of black holes, gravitational lensing and gravitational waves\footnote{While we prepared this contribution, the discovery of gravitational waves
was announced \cite{gwaves}.}. In cosmology, it predicts existence of about
$95 \%$ of additional  new kind of  matter, which makes dark side of the universe. Namely, if GR is the gravity theory for the
universe as a whole and if the universe is homogeneous and isotropic with the flat Friedmann-Lema\^{\i}tre-Robertson-Walker
(FLRW) metric  at the cosmic scale, then it  contains about $68\%$ of {\it dark energy}, $27\%$ of {\it dark  matter},
and only about $5\%$ of {\it visible matter} \cite{planck}.

Despite of some significant phenomenological successes  and many nice theoretical properties, GR is not complete theory
of gravity. For example, attempts to quantize GR lead to the problem of nonrenormalizability. GR  also contains
singularities like the Big Bang and black holes. At the galactic and large cosmic scales GR predicts new forms of matter,
which are not verified in laboratory  conditions and have not so far seen in particle physics.  Hence, there are
many attempts to  modify General relativity. Motivations for its modification   usually  come from  quantum gravity,
string theory, astrophysics and cosmology (for a review, see \cite{clifton,nojiri, faraoni}). We are  mainly
interested in cosmological reasons to modify  Einstein theory of gravity, i.e. to find such extension of GR which
will not contain the Big Bang singularity and  offer another possible description of the universe acceleration and large
velocities in galaxies  instead of  mysterious dark energy and dark matter. It is obvious that physical theory has to be
modified when it contains a singularity. Even if it happened  that dark energy and dark matter really exist it is still
interesting to know is there  a modified gravity which can imitate the same or similar effects. Hence, adequate gravity modification
can reduce role and rate of the  dark matter/energy in the universe.

Any well founded modification of the Einstein theory of gravity has to contain general relativity and to be verified at least on the dynamics of the Solar system. In other words, it has to be a generalization of the general theory of relativity.
 Mathematically, it should be formulated within the pseudo-Riemannian geometry in terms of covariant quantities and take into account equivalence of the inertial and gravitational mass. Consequently, the Ricci scalar $R$ in gravity Lagrangian $\mathcal{L}_g$ of the Einstein-Hilbert action  should  be replaced  by an adequate function which, in general, may contain not only $R$ but also some scalar covariant constructions which are possible in the pseudo-Riemannian geometry. However, we do not know what is here adequate function and
 there are infinitely many  possibilities for its construction. Unfortunately, so far there is no guiding  theoretical principle which could make appropriate choice between all possibilities. In this context the Einstein-Hilbert action is the simplest one, i.e. it can be viewed as realization of the principle of simplicity  in construction of $\mathcal{L}_g$.

One of promising modern  approaches towards more complete theory of gravity
is its nonlocal modification. Motivation for nonlocal modification of general relativity can be found in string theory which is nonlocal theory and contains gravity. We present here a brief review and some new results of nonlocal gravity  with related bounce
cosmological solutions. In particular, we pay special attention to  models in which nonlocality is expressed by an analytic function of the
 d'Alembert
operator $\Box = \frac{1}{\sqrt{-g}} \partial_{\mu}\sqrt{-g} g^{\mu\nu} \partial_{\nu} $ like nonlocality in string theory.  In these models, we are mainly interested in nonsingular bounce
solutions for the cosmic scale factor $a(t)$.

In Sect. 2 we mention a few different approaches to nonlocal modified gravity. Section 3 contains rather general modified action with an analytic nonlocality and with corresponding equations of motion. Cosmological equations for the FLRW metric is  presented in Sect. 4.   Cosmological solutions for constant scalar curvature are considered separately  in Sect. 5. Some new examples of nonlocal models and related Ans\"atze are introduced in Sect. 6. At the and a few  remarks are also noticed.

\section{Nonlocal Modified Gravity}
\label{sec:2}

We consider here nonlocal modified gravity.
Usually a nonlocal modified gravity model contains  an infinite number of spacetime derivatives in the form  of a power series expansion with respect to the d'Alembert
operator $\Box = \frac{1}{\sqrt{-g}} \partial_{\mu}\sqrt{-g} g^{\mu\nu} \partial_{\nu} .$
In this article, we are mainly interested in nonlocality expressed in the form of an analytic function $ \mathcal{F}(\Box)=  \sum_{n =0}^{\infty} f_{n}\Box^{n},$ where coefficients $f_n$ should be determined from various theoretical and phenomenological conditions. Some conditions are related to the absence of tachyons and gosts.

Before to proceed with this analytic nonlocality it is worth to mention some other interesting nonlocal approaches. For approaches containing  $\Box^{-1}$ one can see, e.g., \cite{woodard,woodard-d,woodard1,nojiri1,nojiri2,sasaki,vernov0,vernov1,koivisto,koivisto1} and references therein. For nonlocal gravity with
$\Box^{-1}$ see also \cite{barvinsky,modesto}. Many aspects of nonlocal gravity models have been considered, see e.g.
\cite{capozziello,modesto1,modesto2,moffat,calcagni,maggiore} and references therein.

Our motivation to modify gravity in an analytic nonlocal way comes mainly from string theory, in particular from string field theory (see the very original effort in this direction in \cite{aoriginal}) and $p$-adic string theory
\cite{freund,dragovich,dragovich-d,dragovich-p,vvz}. Since strings are one-dimensional extended objects,  their field theory description contains spacetime nonlocality expressed by some exponential functions of d'Alembert operator $\Box .$ 

  At classical level analytic non-local gravity has proven to alleviate the singularity of the Black-hole type because the Newtonian potential
appears regular (tending to a constant) on a universal basis at the
origin \cite{edholm,biswas1,biswas3}. Also there was significant success in
constructing classically stable solution for the cosmological bounce \cite{biswas1,biswas3+,koshelev,koshelev2,li}.
  
  Analysis of perturbations revealed a natural ability of analytic
non-local gravities to accommodate inflationary models. In particular,
the Starobinsky inflation was studied in details and new predictions for
the observable parameters were made \cite{craps,koshelev4}. Moreover, in the quantum sector
infinite derivative gravity theories improve renormalization, see e.g.
while the unitarity is still preserved \cite{modesto3,modesto4,koshelev4} 
(note that just a local quadratic curvature gravity was proven to be
renormalizable while being non-unitary \cite{stelle}).

\section{Modified GR with Analytical Nonlocality}

To better understand  nonlocal modified gravity itself, we investigate it here without presence of matter. Models of nonlocal gravity which we mainly investigate
are given by the following action
\begin{equation} \label{lag:1}
S =  \int \dx\sqrt{-g}\left(\frac{M_P^2}2R-\Lambda + \frac{\lambda}2P(R)
\FF(\Box)Q(R) \right) ,
\end{equation}
where $R$ is the scalar curvature, $\Lambda$ is the cosmological constant, $
\mathcal{F}(\Box)= \displaystyle \sum_{n =0}^{\infty} f_{n}\Box^{n}$  is an
analytic function  of the d'Alembert-Beltrami operator $\Box =
\nabla^\mu\nabla_\mu$ where $\nabla_\mu$ is the covariant derivative. The
Planck mass $M_P$ is related to the Newtonian constant $G$ as
$M_P^2=\frac1{8 \pi G}$ and $P$,$Q$ are scalar functions of the scalar
curvature. The spacetime dimensionality   $D = 4$ and our signature is $(-,+,+,+)$.
$\lambda$ is a constant and  can be absorbed in the
rescaling of $\FF(\Box)$. However, it is convenient to remain $\lambda$ and recover GR
in the limit  $\lambda\to 0$.

Note that to have physically meaningful  expressions  one should introduce the scale of nonlocality using a new mass
parameter $M$. Then the function $\FF$ would be expanded in Taylor series as
$\mathcal{F}(\Box)= \displaystyle \sum_{n =0}^{\infty} \bar
f_{n}\Box^{n}/M^{2n}$ with all barred constants dimensionless. For simplicity we shall keep $M^2 = 1.$ We shall also see later that analytic function  $ \mathcal{F}(\Box)=  \sum_{n =0}^{\infty} f_{n}\Box^{n},$ has to satisfy some conditions,
in order to escape unphysical degrees of freedom  like ghosts and tachyons, and to have good behavior in quantum sector
(see  \cite{biswas3,biswas4,edholm}).

Varying the action (\ref{lag:1}) by substituting
\be g_{\mu\nu}\to g_{\mu\nu}+h_{\mu\nu}\label{maindelta}\ee to
the linear order in $h_{\mu\nu}$, removing the total derivatives and
integrating from time to time by parts one gets
\begin{equation} \begin{aligned} \label{EOMPQgravity}
\delta S=&\int
\dx\sqrt{-g}\frac{h^{\mu\nu}}2\bigg[-\mathcal{G}_{\mu\nu}\bigg] ,
\end{aligned} \end{equation}
where
\begin{equation} \begin{aligned} \label{EOMPQ}
&\mathcal{G}_{\mu\nu} \equiv M_P^2G_{\mu\nu} + g_{\mu\nu}\Lambda
-\frac{\lambda}2g_{\mu\nu}P\FF(\Box) Q+ \lambda(R_{\mu\nu}-K_{\mu\nu})V
-\frac{\lambda}2\sum_{n=1}^\infty f_n \\ &\times\sum_{l=0}^{n-1}\left(P_\mu^{(l)}
Q_\nu^{(n-l-1)}+P_\nu^{(l)}
Q_\mu^{(n-l-1)}-g_{\mu\nu}(g^{\rho\sigma}P_\rho^{{(l)}}
Q_\sigma^{(n-l-1)}+P^{(l)} Q^{(n-l)})\right) = 0
\end{aligned} \end{equation}
presents equations of motion for gravitational field $g_{\mu\nu}$ in the vacuum.
In (\ref{EOMPQ}) $G_{\mu\nu}=R_{\mu\nu}-\frac12g_{\mu\nu}R$ is the Einstein tensor,
$$
K_{\mu\nu} = \nmu\nnu - g_{\mu\nu} \Box , \qquad V = P_R\FF(\Box)Q + Q_R\FF(\Box)P ,
 $$
 where the subscript $R$ indicates the
derivative w.r.t. $R$ (as many times as it is repeated) and
$$
P^{{(l)}}=\Box^l P,~P_\rho^{{(l)}}=\pd_{\rho}\Box^l P\text{ with the same
for }Q,~P_R,~\dots
$$
In the case of gravity with matter, the full equations of motion are
$\mathcal{G}_{\mu\nu}=T_{\mu\nu},$ where $T_{\mu\nu}$ is the energy-momentum tensor.
Thanks to the integration by parts there is always the symmetry of an
exchange $P\leftrightarrow Q$.

When $\lambda = 0$  in (\ref{EOMPQ}) we recognize  the Einstein's
GR equation with the cosmological constant  $\Lambda$. If $f_n =0$ for $n \geq 1$
then (\ref{EOMPQ}) corresponds to equations of motion of an $f(R)$ theory.


\section{ Cosmological Equations for FLRW  Metric}

We use the FLRW metric
$$ds^2 = - dt^2 + a^2(t)\left(\frac{dr^2}{1-k r^2} + r^2 d\theta^2 +
r^2 \sin^2 \theta d\phi^2\right)$$
and look for some cosmological solutions. In the FLRW
metric the Ricci scalar curvature is
$$R = 6 \left (\frac{\ddot{a}}{a} +
\frac{\dot{a}^{2}}{a^{2}} + \frac{k}{a^{2}}\right )$$
 and $$\Box =
- \partial_t^2  - 3 H \partial_t  ,$$
where $H =
\frac{\dot{a}}{a}$ is the Hubble parameter. We use natural system of units in which
speed of light $c = 1.$

Due to symmetries of the FLRW spacetime,  in (\ref{EOMPQ}) there are only two linearly
independent equations. They are:  trace and $00$, i.e. when indices $\mu = \nu = 0 .$

The trace equation and $00$-equation, respectively,  are
\begin{equation} \begin{aligned} \label{tracePQ}
&M_P^2R-4\Lambda + 2\lambda P \FF(\Box)Q -\lambda(R+3\Box)V  \\
-&{\lambda}\sum_{n=1}^{\infty} f_n
\sum_{l=0}^{n-1}\left(g^{\rho\sigma} \partial_\rho \Box^l P \, \partial_\sigma \Box^{n-l-1} Q+2 \Box^l P \, \Box^{n-l} Q\right)   = 0 ,
\end{aligned} \end{equation}

\begin{equation} \begin{aligned} \label{00PQ}
&M_p^2 G_{00} -\Lambda + \frac{\lambda}{2} P \FF(\Box)Q + \lambda (R_{00} - \nabla_0 \nabla_0 - \Box) V
-\frac{\lambda}{2} \sum_{n=1}^\infty f_n  \\ &\times \sum_{l=1}^{n-1} \left( 2 \partial_0 \Box^l P \, \partial_0 \Box^{n-l-1} Q +g^{\rho\sigma}
\partial_\rho \Box^l P \, \partial_\sigma \Box^{n-l-1} Q  + \Box^l P \, \Box^{n-l} Q \right) = 0 .
\end{aligned} \end{equation}

\section{Cosmological Solutions for Constant Scalar Curvature $R$}

When $R$ is a constant then $P$ and $Q$ are also some constants and we have that $\Box R = 0 ,$  $\FF(\Box)=f_0 .$ The corresponding equations of motion \eqref{tracePQ} and \eqref{00PQ} contain solutions as in the local case. However,  metric perturbations at the background $R = const.$  can give nontrivial cosmic structure due to nonlocality.

Let $R = R_0 = constant \neq 0. $  Then
\begin{equation} \label{eq6}
6\Big(\frac{\ddot a}{a} + \big(\frac{\dot a}{a}\big)^2 +
\frac{k}{a^2}\Big)= R_0.
\end{equation}

The change of variable $b(t) = a^2(t)$ transforms \eqref{eq6} into equation
\begin{equation} \label{eq7}
3 \ddot b - R_0 b = -6k.
\end{equation}

Depending on the sign of $R_0$, the following solutions of equation \eqref{eq7} are
\begin{equation} \begin{aligned} \label{11.03.13:1}
 & b(t) = \frac{6k}{R_0} + \sigma e^{\sqrt{ \frac {R_0}3}\, t} + \tau e^{- \sqrt{ \frac {R_0}3}\, t}, \quad R_0 > 0 , \\
& b(t) = \frac{6k}{R_0} + \sigma \cos {\sqrt{ \frac {-R_0}3}\, t} + \tau \sin {\sqrt{ \frac {-R_0}3}\, t}, \quad R_0 < 0 ,
\end{aligned}\end{equation}
where $\sigma$ and $\tau$ are some constant coefficients.

Substitution $R = R_0 $ into equations of motion \eqref{tracePQ} and \eqref{00PQ} yields, respectively,
\begin{align}  \label{11.03.13:2}
&M_p^2R_0 -4\Lambda +2\lambda P f_0 Q-\lambda R_0V_0 = 0 , \\
&M_p^2 G_{00} -\Lambda + \frac{\lambda}{2}P f_0 Q + \lambda R_{00} V_0 = 0 ,\label{11.03.13:2a}
\end{align}
where $V_0 = f_0 (P_R Q + Q_R P)|_{R=R_0}$ and $G_{00} = R_{00} + \frac{R_0}{2} .$

Combining  equations \eqref{11.03.13:2} and \eqref{11.03.13:2a} one obtains
\begin{align}\label{eq1:1}
&M_p^2R_0 -4\Lambda +2\lambda P f_0 Q-\lambda R_0V_0 = 0, \\
&4 R_{00} +  R_0 = 0. \label{eq1:2}
\end{align}

Equation \eqref{eq1:1} connects some parameters of the nonlocal model \eqref{lag:1} in the algebraic form with respect to $R_0$, while
 \eqref{eq1:2} implies a condition on the parameters $\sigma, \tau, k$
and  $R_0$ in solutions  \eqref{11.03.13:1}.
Namely,   $R_{00}$ is related to function $b(t)$ as
\begin{equation}  \label{eq2:1}
R_{00} = - \frac{3\ddot a}a= \frac{3}{4}\frac{(\dot b)^2 - 2b \ddot b}{b^2}.
\end{equation}
Replacing $R_{00}$ in \eqref{eq1:2} by  \eqref{eq2:1} and using different solutions for $b(t)$ in \eqref{11.03.13:1} we obtain
\begin{equation} \begin{aligned} \label{11.03.13:3}
 & 9k^2 = R_0^2\sigma\tau, \, \qquad R_0 > 0, \\
& 36k^2 = R_0^2(\sigma^2+ \tau^2), \, \qquad  R_0 < 0.
\end{aligned} \end{equation}


\subsection{Case: \, $R_0 > 0$}

\begin{itemize}
 \item
 Let $k=0$. From $9k^2 = R_0^2 \sigma \tau$ follows that at least one of $\sigma$ and $\tau$ has to be zero. Thus there is possibility for
an exponential solution for $a(t)$ and $a(t) = 0.$  Taking $\tau =0$  and $\sigma = a_0^2$
 one has
\begin{align}
 b(t) =  a_0^2\,  e^{\sqrt{ \frac {R_0}3}\, t}.
\end{align}

\item
If  $k=+1$ one can find $\varphi$ such that
$\sigma+\tau = \frac{6}{R_0} \cosh \varphi$ and $\sigma-\tau =
\frac{6}{R_0} \sinh\varphi$. Moreover, we obtain
\begin{equation} \begin{aligned}
b(t) &= \frac{12}{R_0} \cosh^2 \frac 12 \Big(\sqrt{\frac{R_0}{3}} t + \varphi\Big), \\
a(t) &= \sqrt{\frac{12}{R_0}} \cosh \frac 12 \Big(\sqrt{\frac{R_0}{3}} t + \varphi\Big).
\end{aligned}\end{equation}
\item
If  $k=-1$ one can transform $b(t)$  and $a(t)$ to
\begin{equation} \begin{aligned}
b(t) &= \frac{12}{R_0} \sinh^2 \frac 12 \Big(\sqrt{\frac{R_0}{3}} t + \varphi\Big), \\
a(t) &= \sqrt{\frac{12}{R_0}} \Big|\sinh \frac 12 \Big(\sqrt{\frac{R_0}{3}} t + \varphi\Big)\Big|.
\end{aligned}\end{equation}
\end{itemize}

\subsubsection{Case: $R= 12 \gamma^{2}$}

This is a special case of $R_0$, which simplifies the above expressions and yields de Sitter-like cosmological solutions.

\begin{itemize}
\item $k = 0$:
\begin{align}
 b(t) =  a_0^2\,  e^{2 \,\gamma\, t} ,  \quad  a(t) =  a_0\,  e^{\gamma\, t} .
\end{align}

\item  $k=+1$:
\begin{equation} \begin{aligned}
b(t) &= \frac{1}{\gamma^2} \cosh^2  \Big(\gamma\, t + \frac{\varphi}{2}\Big), \\
a(t) &= \frac{1}{|\gamma|} \cosh \Big(\gamma \, t + \frac{\varphi}{2}\Big).
\end{aligned}\end{equation}
\item $k=-1$:
\begin{equation} \begin{aligned}
b(t) &= \frac{1}{\gamma^2} \sinh^2 \Big( \gamma \, t + \frac{\varphi}{2}\Big), \\
a(t) &= \frac{1}{|\gamma|} \Big|\sinh \Big(\gamma \, t + \frac{\varphi}{2}\Big)\Big|.
\end{aligned}\end{equation}
\end{itemize}

\subsection{Case: \, $R_0 < 0$}

\begin{itemize}

\item
When $k=0$ then $\sigma = \tau = 0,$ and consequently $b(t) = 0.$

\item
If $k=-1$ one can define $\varphi$ by
$\sigma = \frac{-6}{R_0} \cos \varphi$ and $\tau = \frac{-6}{R_0}
\sin \varphi$, and rewrite  $b(t)$ and $a(t)$ as
\begin{equation}  \label{eq13}   \begin{aligned}
b(t) &= \frac{-12}{R_0} \cos^2 \frac 12 \Big(\sqrt{-\frac{R_0}{3}} t - \varphi\Big), \\
a(t) &= \sqrt{\frac{-12}{R_0}} \Big|\cos \frac 12 (\sqrt{-\frac{R_0}{3}} t - \varphi)\Big|.
\end{aligned}\end{equation}

\item
In the last case $k=+1$, by the same procedure as for $k = -1,$ one can transform $b(t)$ to expression
\begin{equation} \begin{aligned}
b(t) &= \frac{12}{R_0} \sin^2 \frac 12 \Big(\sqrt{-\frac{R_0}{3}} t - \varphi\Big),
\end{aligned}\end{equation}
which is not positive and hence yields no solution.
\end{itemize}

\subsection{Case: \, $R_0 = 0$}

The case $R_0 = 0$ can be considered as limit of  $R_0 \to 0$ in both cases $R_0 > 0$ and $R_0 < 0.$
When $R_0 > 0$ there is condition $9k^2 = R_0^2 \sigma\tau $ in \eqref{11.03.13:3}. From this condition, $R_0 \to 0$ implies $k = 0$ and arbitrary values of constants $\sigma$ and $\tau$. The  same conclusion obtains when $R_0 < 0$ with condition $36k^2 = R_0^2(\sigma^2+ \tau^2)$.     In both these cases there is  Minkowski solution with $b(t) = constant >0$ and consequently $a(t) = constant > 0,$ see \eqref{11.03.13:1}.

\section{Some Models and Related Ans\"atze for Cosmological Solutions}

\subsection{Nonlocal Gravity Model Quadratic in $R$}

Nonlocal gravity model which is quadratic in $R$ was given by the action \cite{biswas1,biswas2}
\begin{equation}
S =  \int d^{4}x \sqrt{-g}\Big(\frac{R - 2 \Lambda}{16 \pi G} +  R
\mathcal{F}(\Box) R  \Big).             \label{eq:3.1}
\end{equation}
This model is important because it is ghost free and has some nonsingular bounce solutions, which can be regarded as a solution of
the Big Bang cosmological singularity problem.

The corresponding equations of motion can be easily obtained from \eqref{tracePQ}  and \eqref{00PQ}.
To evaluate related equations of motion, the following Ans\"atze  were used:
\begin{itemize}
\item   Linear Ansatz: $\Box R = r R + s, $ where $r$ and $s$ are constants.
\item   Quadratic Ansatz:  $\Box R = q R^2, $ where $q$ is a constant.
\item   Qubic Ansatz:  $\Box R = C R^3, $ where $C$ is a constant.
\item    Ansatz $\Box^n R = c_n R^{n+1}, \,\, n\geq 1, $ where $c_n$ are constants.
\end{itemize}
These Ans\"atze make some constraints on possible solutions, but
simplify formalism to find a particular solution (see \cite{dragovich1} and references therein).

\subsubsection{Linear Ansatz and Nonsingular Bounce Cosmological Solutions}

Using Ansatz  $\Box R = r R + s$ a few nonsingular bounce solutions for  the scale factor are found:
$a(t) = a_0 \cosh{\left(\sqrt\frac{\Lambda}{3}t\right)}$ (see \cite{biswas1,biswas2}),
 $\, a(t) = a_0 e^{\frac{1}{2}\sqrt{\frac{\Lambda}{3}}t^2}$ (see \cite{koshelev,koshelev0}) and $a(t) = a_0  (\sigma  e^{\lambda t} + \tau e^{-\lambda t} )$
 \cite{dragovich2}.
The first two consequences of this Ansatz are
\begin{equation}\label{nth degree}
\Box^{n} R = r^{n}(R +\frac sr ) , \, \, n\geq 1 , \,  \qquad \, \mathcal{F}(\Box)
R = \mathcal{F}(r) R + \frac sr(\FF(r)-f_0) ,
\end{equation}
which considerably simplify nonlocal term.

Generalization of the above quadratic model in the form of nonlocal term  $R^{p} \mathcal{F}(\Box) R^q ,$ where $p$ and $q$ are some natural numbers,
was recently considered in \cite{dimitrijevic1}. Here cosmological solution for the  scale factor has the form $a(t) = a_o \, e^{-\gamma\, t^2}.$

\subsection{Gravity Model with Nonlocal Term $ R^{-1} \mathcal{F}(\Box) R $}

This model was introduced in \cite{dragovich3} and its action may be written in the form
\begin{equation} \label{eq-3.2.1}
S =  \int d^{4}x \sqrt{-g}\Big(\frac{R}{16 \pi G} +   R^{-1} \mathcal{F}(\Box) R  \Big),
\end{equation}
where $\mathcal{F}(\Box) = \sum_{n=0}^\infty f_n \Box^n $ and  $f_0 = -\frac{\Lambda}{8\pi G}$  plays role of the
cosmological constant.

The nonlocal term $R^{-1} \mathcal{F}(\Box) R$ in \eqref{eq-3.2.1} is invariant under transformation $R \to C R.$  This nonlocal term does not
depend on the magnitude of scalar curvature $R,$ but on its spacetime dependence, and in the FLRW case is relevant  only  dependence of $R$ on time $t$.
 Term $f_0 = -\frac{\Lambda}{8\pi G}$ is completely determined by the cosmological constant $\Lambda ,$ which according to $\Lambda CDM$ model is small and positive energy density of the vacuum. Coefficients $f_i , \,\, i \in \mathbb{N}$ can be estimated from other conditions, including agreement with dynamics   the Solar system. In comparison to the model quadratic in $R$ \eqref{eq:3.1}, complete Lagrangian of this model remains to be linear in $R $ and in such sense is simpler nonlocal modification than \eqref{eq:3.1}.

In this model   are also used the above Ans\"atze. Especially quadratic Ansatz  $\Box R = q R^2, $ where $q$ is a constant, is effective to consider power-law cosmological solutions, see \cite{dragovich3,dragovich4,dragovich5,dragovich6}.

\subsection{Some New Models and Ans\"atze}

It is worth  to consider some particular examples  of action \eqref{lag:1} when $P = Q = (R + R_0)^m ,$ i.e.
\begin{equation}
S = \int \Big(\frac{1}{16 \pi G}R   -\Lambda +
\frac{\lambda}{2} (R + R_0)^m \mathcal{F}(\Box)(R + R_0)^m \Big)\sqrt{-g} \; \dx ,
\end{equation}
where $ R_0 \in \mathbb{R}, \, m  \in \mathbb{Q},$ and which have scale factor solution as
\begin{equation} \label{eq-6.3.1}
a(t) = A \; t^{n} e^{\gamma\, t^{2}},  \quad \gamma \in \mathbb{R}.
\end{equation}

To this end we consider the Ansatz
\begin{align} \label{eq-6.3.2}
\Box (R+R_{0})^{m} = p (R+R_{0})^{m},
\end{align}
where  $p$ is a constant and $\Box$ is the d'Alembert operator in FLRW metric.

>From Ansatz \eqref{eq-6.3.2} and scalar curvature $R$ for $k=0 ,$ we get the following system of equations:
\begin{equation}\begin{aligned} \label{eq-6.3.3}
 &72 m (1 + 2 m - 3 n) n^2 (-1 + 2 n)^2=0, \\
&36 n (-1 + 2 n) (-n p + 2 n^2 p + m R_{0} - m n R_{0} + 12 m \gamma + 48 m n \gamma - 72 m n^2 \gamma)=0, \\
&12 n (-1 + 2 n) (p R_{0} + 12 p \gamma + 48 n p \gamma - 6 m R_{0} \gamma + 312 m \gamma^2 - 192 m^2 \gamma^2 -288 m n \gamma^2)=0,\\
&p R_{0}^2 + 24 p R_{0} \gamma + 96 n p R_{0} \gamma + 144 p \gamma^2 +576 n p \gamma^2 +3456 n^2 p \gamma^2 + 96 m R_{0} \gamma^2 +
   288 m n R_{0} \gamma^2  \\
&+ 1152 m \gamma^3 + 8064 m n \gamma^3 + 13824 m n^2 \gamma^3=0, \\
&96 \gamma^2 (p R_{0} + 12 p \gamma + 48 n p \gamma + 6 m R_{0} \gamma + 24 m \gamma^2 + 96 m^2 \gamma^2 + 432 m n \gamma^2)=0,\\
&2304 \gamma^4 (p + 12 m \gamma)=0.
\end{aligned}\end{equation}

System of equations \eqref{eq-6.3.3} has $5$ solutions:
\begin{enumerate}
\item $p= -12 m \gamma$, $n=0$, $R_{0}=-12 \gamma$, $m=\frac{1}{2}$
\item $p= -12 m \gamma$, $n=\frac{2m+1}{3}$, $R_{0}=-28 \gamma$, $m=\frac{1}{2}$
\item $p= -12 m \gamma$, $n=0$, $R_{0}=-4 \gamma$, $m=1$
\item $p= -12 m \gamma$, $n=\frac{1}{2}$, $R_{0}=-16 \gamma$, $m=1$
\item $p= -12 m \gamma$, $n=\frac{1}{2}$, $R_{0}=-36 \gamma$, $m=-\frac{1}{4}$
\end{enumerate}

We shall now shortly consider each of the above cases.

\subsubsection{Case 1: $a(t) =A \, e^{\gamma\, t^{2}} , $   $m = \frac{1}{2}$}

Here  Ansatz is $\Box \sqrt{R+R_{0}} = p \sqrt{R+R_{0}}$, where $R_{0}=-12 \gamma ,$  $p= -6 \gamma$ and $\gamma$ is a parameter.
The scale factor  is $a(t) = A \, e^{\gamma\, t^{2}} .$

The first consequences of this  Ansatz are
\begin{align*}
\Box^{\ell} \sqrt{R+R_{0}} &= p^{\ell}\sqrt{R+R_{0}} , \; \, \ell\geq 0 , \\
\mathcal{F}(\Box) \sqrt{R+R_{0}} &= \mathcal{F}(p) \sqrt{R+R_{0}} , \\
R(t)&= 12 \gamma ( 1+ 4 \gamma\, t^{2}).
\end{align*}

Relevant action is
\begin{equation}
S = \int \Big(\frac{1}{16 \pi G}R   -\Lambda +
\frac{\lambda}{2} \sqrt{R-12 \gamma} \mathcal{F}(\Box)\sqrt{R-12 \gamma} \Big)\sqrt{-g} \; \dx.
\end{equation}

Equations of motion follow  from \eqref{tracePQ} and \eqref{00PQ}, where $P= Q = \sqrt{R - 12 \gamma}.$
Straightforward calculation gives cosmological solution $a(t) =A\, e^{\gamma\, t^{2}}$ with conditions:

$$
\mathcal{F}(p) = \frac{\gamma - 4\pi G \Lambda}{16 \gamma \pi G \lambda},   \quad
\mathcal{F}'(p) = \frac{4\pi G \Lambda -3 \gamma}{192 \gamma^{2} \pi G \lambda} , \quad p = - 6 \gamma .
$$

\subsubsection{Case 2:  $a(t) = A\, t^{2/3} e^{\gamma \, t^{2}} ,$  $m = \frac{1}{2}$}
In this case the Ansatz is $\Box \sqrt{R+R_{0}} = p \sqrt{R+R_{0}}$, where $R_{0}$ and $p$ are real constants.

The first consequences of this Ansatz are
\begin{align*}
\Box^{\ell} \sqrt{R+R_{0}} &= p^{\ell}\sqrt{R+R_{0}} , \; \, \ell \geq 0 ,\\
\mathcal{F}(\Box) \sqrt{R+R_{0}} &= \mathcal{F}(p) \sqrt{R+R_{0}}.
\end{align*}
For scale factor $ a(t) = A\, t^{2/3} e^{\gamma\, t^{2}}$
 the Ansatz $\Box \sqrt{R+R_{0}} = p \sqrt{R+R_{0}}$ is satisfied if and only if $ R_{0}=-28 \gamma$ and $p=-6 \gamma$.

Direct calculation shows that
\begin{align*} \label{R-R'}
 R(t)&= 44 \gamma+ \frac{4}{3}t^{-2} + 48 \gamma^{2} t^{2}, \\
\Box^{\ell} \sqrt{R-28 \gamma} &= (-6 \gamma)^{\ell} \sqrt{R-28 \gamma}, \; \; \ell \geq 0 ,  \\
\mathcal{F}(\Box) \sqrt{R-28 \gamma} &= \mathcal{F}(-6 \gamma)\sqrt{R-28 \gamma},\\
 \dot R &= 96 \gamma^{2}t- \frac{8}{3}t^{-3}.
\end{align*}

The related action is
\begin{equation}
S = \int \Big(\frac{1}{16 \pi G}R - \Lambda +
\frac{\lambda}{2} \sqrt{R-28 \gamma} \mathcal{F}(\Box)\sqrt{R-28 \gamma}\Big)\sqrt{-g} \; \dx.
\end{equation}
The corresponding trace and $00$ equations of motion are satisfied
under conditions:
$$
   \mathcal{F}(p) = -\frac{1}{8 \pi G \lambda}, \quad \mathcal{F}'(p) = 0, \quad \gamma  = \frac{4 }{7 }  \pi G \Lambda , \quad p = - 6 \gamma .
$$

\subsubsection{Case 3:  $a(t) = A\, e^{\gamma \, t^{2}} ,$  $m=1$}

In this case $\Box (R - 4 \gamma) = - 12 \gamma (R - 4\gamma) $, what is an example of already above considered linear Ansatz.
The corresponding action is
\begin{equation}
S = \int \Big(\frac{1}{16 \pi G}R - \Lambda +
\frac{\lambda}{2} (R-4 \gamma) \mathcal{F}(\Box) (R- 4 \gamma)\Big)\sqrt{-g} \; \dx.
\end{equation}
Equations of motion have cosmological solution $a(t) = A\,  e^{\gamma \, t^{2}} $ under conditions:

$$ \mathcal{F}(p)= - \frac{1}{512\pi G \lambda \gamma}, \quad \mathcal{F}'(p)=0, \quad p= -12 \gamma , \quad \gamma =  8 \pi G \Lambda .$$

\subsubsection{Case 4:  $a(t) = A\, \sqrt{t} e^{\gamma \, t^{2}} ,$  $m=1$}

This case is quite similar to the previous one. Now Ansatz is $\Box (R - 16 \gamma) = - 12 \gamma (R - 16\gamma) $ and action

\begin{equation}
S = \int \Big(\frac{1}{16 \pi G}R - \Lambda +
\frac{\lambda}{2} (R-16 \gamma) \mathcal{F}(\Box) (R- 16 \gamma)\Big)\sqrt{-g} \; \dx.
\end{equation}
Scale factor $a(t) = A\, \sqrt{t} e^{\gamma \, t^{2}}$ is solution of equations of motion if the following conditions are satisfied:

$$\mathcal{F}(p)= - \frac{1}{320\pi G \lambda \gamma}, \quad \mathcal{F}'(p)=0 , \quad  p= -12 \gamma , \quad \gamma =  8 \pi G \Lambda .$$

\subsubsection{Case 5: $a(t) = A\, \sqrt{t}\, e^{\gamma \, t^{2}} ,$  $m=-\frac{1}{4}$}
According to the Ansatz, in this case $p= 3 \gamma$, $n=\frac{1}{2}$, $R_{0}=-36 \gamma .$  However the action
\begin{equation}
S = \int \Big(\frac{1}{16 \pi G}R - \Lambda +
\frac{\lambda}{2} \sqrt{R-36 \gamma} \mathcal{F}(\Box)\sqrt{R-36 \gamma}\Big)\sqrt{-g} \; \dx.
\end{equation}
has no solution $a(t) = A\, \sqrt{t} \, e^{\gamma \, t^{2}}$ for the Ansatz $\Box (R+R_{0})^{m} = p (R+R_{0})^{m}, \, \, m=-\frac{1}{4}.$

\section{Concluding Remarks}

In this paper we presented a brief review of nonlocal modified gravity, where nonlocality is realized by an analytic function of the d'Alembert operator
$\Box$. Considered models are presented by actions, their equations of motion, related Ans\"atze and some cosmological solutions for the scale factor $a(t)$. A few new models are introduced, and they deserve to be further investigated, especially Case 1 and Case 2 in section 6.

Many details on \eqref{lag:1} and its extended  versions can be found in
\cite{biswas3,biswas4,biswas3+,koshelev0,koshelev1,koshelev2}. Perturbations  and physical excitations of the equations of motion of action \eqref{eq:3.1} around the de Sitter background are considered in \cite{dragovich7} and \cite{dragovich8}, respectively. As some recent developments in nonlocal modified gravity, see \cite{golovnev,chicone,cusin,edholm,koshelev4,zhang}.

Notice that nonlocal cosmology is related also to cosmological models in which matter sector contains nonlocality (see, e.g.
 \cite{arefeva,av2,calcagni1,barnaby,koshelev-v,dragovich,dragovich-d}).
 String field theory and $p$-adic string theory models have played significant role in motivation and construction of such models. One particular aspect in which non-local models prove important is the ability to resolve the Null Energy Condition obstacle \cite{av1} common to many models of generalized gravity. In short, that is an ability to construct a healthy model which has sum of energy and pressure of the matter positive and thereby avoids ghosts in the spectrum alongside with a nonsingular space-time structure \cite{raych}.

Nonsingular bounce cosmological solutions are very important (as reviews on bouncing cosmology, see e.g.  \cite{novello,brandenberger})
and their progress in nonlocal gravity may be a further step towards cosmology of the cyclic universe \cite{steinhardt}.

\begin{acknowledgement}
Work on this paper was partially supported by Ministry of Education, Science and Technological Development of the Republic of Serbia, grant No 174012.
B.D. thanks Prof. Vladimir Dobrev for invitation to participate and give a talk on nonlocal gravity, as well as for hospitality, at the XI International Workshop ``Lie Theory
and its Applications in Physics'', 15--21 June 2015, Varna, Bulgaria. B.D. also thanks a support of the ICTP - SEENET-MTP project PRJ-09 ``Cosmology and Strings'' during preparation of this article. AK is supported by the FCT Portugal fellowship SFRH/BPD/105212/2014 and
in part by FCT Portugal grant UID/MAT/00212/2013 and by RFBR grant
14-01-00707.
\end{acknowledgement}


\begin{thebibliography}{99.}%

\bibitem{gwaves}  Abbott, B. P., et al., (LIGO Scientific Collaboration and Virgo Collaboration): Observation of gravitational waves from a binary black hole merger. Phys. Rev. Lett. \textbf{116}, 061102 (2016)

\bibitem{planck}  Ade, P. A. R., Aghanim, N., Armitage-Caplan, C., et al. (Planck Collaboration): Planck 2013 results. XVI.
Cosmological parameters.  [arXiv:1303.5076v3]

\bibitem{aoriginal}
Aref'eva, I.Ya., Nonlocal string tachyon as a model for cosmological dark energy.
  AIP Conf. Proc.  {\bf 826}, 301 (2006). [astro-ph/0410443].

\bibitem{arefeva}   Aref'eva, I.Ya.,  Joukovskaya, L.V.,   Vernov, S.Yu.: Bouncing and accelerating
solutions in nonlocal stringy models.  JHEP \textbf{0707}, 087 (2007)
 [hep-th/0701184]

\bibitem{av1}
	Aref'eva, I.Ya., Volovich, I.V., On the null energy condition and cosmology.  Theor. Math. Phys.  {\bf 155}, 503 (2008).
	[hep-th/0612098].

 \bibitem{av2}
Aref'eva, I.Ya., Volovich, I.V.,
Cosmological Daemon. JHEP {\bf 1108}, 102 (2011).


\bibitem{barnaby}  Barnaby, N.,  Biswas, T., Cline,  J.M.: $p$-Adic inflation. JHEP \textbf{0704}, 056 (2007)
 [hep-th/0612230]

\bibitem{barvinsky}  Barvinsky, A.O.: Dark energy and dark matter from nonlocal ghost-free gravity theory. Phys. Lett. B \textbf{710}, 12--16 (2012).
 	[arXiv:1107.1463 [hep-th]]

\bibitem{biswas3} Biswas, T.,  Gerwick, E., Koivisto, T., Mazumdar, A.: Towards singularity and ghost free theories
of gravity.  Phys. Rev. Lett. \textbf{108}, 031101 (2012). [arXiv:1110.5249v2 [gr-qc]]

\bibitem{biswas4} Biswas, T., Conroy, A., Koshelev, A.S., Mazumdar, A.: Generalized gost-free quadratic curvature gravity.
[arXiv:1308.2319 [hep-th]]

\bibitem{biswas1}  Biswas, T.,  Mazumdar, A.,  Siegel, W: Bouncing universes in string-inspired gravity. J. Cosmology
Astropart. Phys. \textbf{0603},  009 (2006). [arXiv:hep-th/0508194]

\bibitem{biswas2}  Biswas, T.,   Koivisto, T.,  Mazumdar, A.: Towards  a resolution of the cosmological
singularity in non-local higher derivative theories of gravity.
J. Cosmology Astropart. Phys. \textbf{1011}, 008 (2010). [arXiv:1005.0590v2 [hep-th]].

\bibitem{biswas3+}  Biswas, T.,  Koshelev, A.S., Mazumdar, A., Vernov, S.Yu.: Stable bounce and inflation in non-local higher
derivative cosmology.   J. Cosmology Astropart. Phys. \textbf{08}, 024 (2012). [arXiv:1206.6374 [astro-ph.CO]]

\bibitem{brandenberger} Brandenberger, R.H.:   The matter bounce alternative to inflationary cosmology.  	[arXiv:1206.4196 [astro-ph.CO]]

\bibitem{freund} Brekke, L.,   Freund, P.G.O.: $p$-Adic numbers in physics. Phys. Rep. \textbf{233},  1--66 (1993).

\bibitem{modesto1} Briscese, F., Marciano, A.,  Modesto, L.,  Saridakis, E.N.: Inflation in (super-)renormalizable gravity. Phys. Rev. D \textbf{87},
083507 (2013). [arXiv:1212.3611v2 [hep-th]]

\bibitem{modesto2} Calcagni, G., Modesto, L., Nicolini, P.: Super-accelerting bouncing cosmology in assymptotically-free non-local gravity.
 	[arXiv:1306.5332 [gr-qc]]

\bibitem{calcagni}     Calcagni, G., Nardelli, G.: Nonlocal gravity and the diffusion equation.
Phys. Rev. D \textbf{82}, 123518 (2010).  	[arXiv:1004.5144 [hep-th]]

\bibitem{calcagni1} Calcagni, G.,   Montobbio, M.,   Nardelli, G.: A route to nonlocal cosmology.
Phys. Rev. D \textbf{76}, 126001 (2007).  [arXiv:0705.3043v3 [hep-th]]

\bibitem{capozziello} Capozziello S., Elizalde E., Nojiri S., Odintsov S.D.: Accelerating cosmologies from non-local higher-derivative
gravity. Phys. Lett. B \textbf{671}, 193 (2009).  [arXiv:0809.1535]

\bibitem{chicone} Chicone, C.,  Mashhoon, B.:  Nonlocal gravity in the solar system.  Class. Quantum Grav. \textbf{33}, 075005 (2016).
[arXiv:1508.01508 [gr-qc]]

\bibitem{clifton}  Clifton,T.,  Ferreira, P.G.,  Padilla, A.,   Skordis, C.:  Modified gravity and cosmology.
Phys. Rep. \textbf{513}, 1--189 (2012).  [arXiv:1106.2476v2 [astro-ph.CO]]

\bibitem{raych} 
  Conroy, A., Koshelev, A. S., Mazumdar, A.,
  Geodesic completeness and homogeneity condition for cosmic inflation.
  Phys. Rev. D {\bf 90}, no. 12, 123525 (2014).
  [arXiv:1408.6205 [gr-qc]].

\bibitem{craps}  Craps, B., de Jonckheere, T., Koshelev, A.S.:  Cosmological perturbations in non-local
higher-derivative gravity. [arXiv:1407.4982 [hep-th]]

\bibitem{cusin} Cusin, G.,  Foffa, S.,  Maggiore, M.,  Michele Mancarella, M.: Conformal symmetry and nonlinear extensions of nonlocal gravity.
Phys. Rev. D \textbf{93}, 083008 (2016). [arXiv:1602.01078 [hep-th]]

\bibitem{woodard-d}  Deffayet, C.,  Woodard, R.P.:  Reconstructing the distortion function for nonlocal cosmology.  JCAP \textbf{0908}, 023 (2009).
 [arXiv:0904.0961 [gr-qc]]

\bibitem{woodard} Deser, S., Woodard, R.P.: Nonlocal cosmology. Phys. Rev. Lett. \textbf{99}, 111301 (2007). [ arXiv:0706.2151 [astro-ph]]

\bibitem{dimitrijevic1}  Dimitrijevic, I.: Cosmological solutions in modified gravity with monomial nonlocality.  Appl. Math.  Comput.  195--203 (2016). [arXiv:1604.06824 [gr-qc]]

\bibitem{dragovich1}  Dimitrijevic, I., Dragovich, B., Grujic J., Rakic, Z.: On modified gravity. Springer Proceedings in Mathematics $\&$
Statistics \textbf{36}, 251--259 (2013). [arXiv:1202.2352 [hep-th]]

\bibitem{dragovich2} Dimitrijevic, I., Dragovich, B., Grujic J., Rakic, Z.: New cosmological solutions in nonlocal modified gravity.
Rom. Journ. Phys. \textbf{58}  (5-6), 550--559 (2013). [arXiv:1302.2794 [gr-qc]]

\bibitem{dragovich3} Dimitrijevic, I., Dragovich, B., Grujic J., Rakic, Z.: A new model of nonlocal modified gravity.
Publications de l'Institut Mathematique \textbf{94} (108), 187--196 (2013)

\bibitem{dragovich4} Dimitrijevic, I., Dragovich, B., Grujic J., Rakic, Z.: Some pawer-law cosmological solutions in nonlocal modified gravity.
Springer Proceedings in Mathematics $\&$ Statistics \textbf{111}, 241--250 (2014)

\bibitem{dragovich6} Dimitrijevic, I., Dragovich, B., Grujic J., Rakic, Z.: Some cosmological solutions of a nonlocal modified gravity.  Filomat \text{29} (3),
619--628. arXiv:1508.05583 [hep-th]

\bibitem{dragovich7} Dimitrijevic, I., Dragovich, B., Grujic J., Koshelev A. S., Rakic, Z.: Cosmology of modified gravity with a non-local $f(R)$.
	arXiv:1509.04254 [hep-th]

\bibitem{dragovich8} Dimitrijevic, I., Dragovich, B., Grujic J., Koshelev A. S., Rakic, Z.: Paper in preparation.

\bibitem{maggiore}  Dirian, Y., Foffa, S., Khosravi, N., Kunz, M.,  Maggiore, M.: Cosmological perturbations and structure formation
in nonlocal infrared modifications of general relativity.  	[arXiv:1403.6068 [astro-ph.CO]]

\bibitem{dragovich5}  Dragovich, B.: On nonlocal modified gravity and cosmology. Springer Proceedings in Mathematics $\&$ Statistics \textbf{111}, 251--262 (2014). arXiv:1508.06584 [gr-qc]

\bibitem{dragovich}  Dragovich, B.: Nonlocal dynamics of $p$-adic  strings.
Theor. Math. Phys. \textbf{164} (3), 1151--115  (2010). [arXiv:1011.0912v1 [hep-th]]

\bibitem{dragovich-d}  Dragovich, B.: Towards $p$-adic matter in the universe. Springer Proceedings in Mathematics and
Statistics \textbf{36}, 13--24 (2013). [arXiv:1205.4409 [hep-th]]

\bibitem{dragovich-p} Dragovich, B., Khrennikov, A. Yu., Kozyrev, S. V., Volovich, I. V.: On p-adic mathematical physics. p-Adic Numbers Ultrametric Anal. Appl.
\textbf{1} (1), 1--17 (2009). 	[arXiv:0904.4205 [math-ph]]

\bibitem{edholm} Edholm, J.,  Koshelev, A. S.,  Mazumdar, A.: Universality of testing ghost-free gravity.  [arXiv:1604.01989 [gr-qc]]

\bibitem{vernov0} Elizalde, E., Pozdeeva, E.O., Vernov, S.Yu.: Stability of de Sitter solutions in non-local cosmological models.  PoS(QFTHEP2011) 038, (2012).
[arXiv:1202.0178 [gr-qc]]

\bibitem{vernov1} Elizalde, E., Pozdeeva, E.O., Vernov, S.Yu., Zhang, Y.: Cosmological solutions of a nonlocal model with a perfect fluid.
  J. Cosmology Astropart. Phys.   \textbf{1307}, 034 (2013). [arXiv:1302.4330v2 [hep-th]]

\bibitem{golovnev}  Golovnev, A., Koivisto, T.,  Sandstad, M.:  Effectively nonlocal metric-affine gravity.  Phys. Rev. D \textbf{93}, 064081 (2016).  [arXiv:1509.06552v2 [gr-qc]]

\bibitem{nojiri2}
Jhingan, S.,  Nojiri, S.,  Odintsov, S.D.,  Sami,  Thongkool M.I.,   Zerbini, S.:
Phantom and non-phantom dark energy: The Cosmological relevance of non-locally corrected gravity.  Phys. Lett. B \textbf{663}, 424-428 (2008).
[arXiv:0803.2613 [hep-th]]

\bibitem{koivisto}  Koivisto, T.S.: Dynamics of nonlocal cosmology.   Phys. Rev. D \textbf{77}, 123513 (2008).   [arXiv:0803.3399 [gr-qc]]

\bibitem{koivisto1}  Koivisto, T.S.: Newtonian limit of nonlocal cosmology.  Phys. Rev. D \textbf{78}, 123505  (2008).   [arXiv:0807.3778 [gr-qc]]

\bibitem{koshelev}  Koshelev, A.S.,  Vernov, S.Yu.: On bouncing solutions in non-local gravity.
[arXiv:1202.1289v1 [hep-th]]

\bibitem{koshelev0} Koshelev, A.S.,  Vernov, S.Yu.: Cosmological solutions in nonlocal models.
[arXiv:1406.5887v1 [gr-qc]]

\bibitem{koshelev1}  Koshelev, A.S.: Modified non-local gravity. [arXiv:1112.6410v1 [hep-th]]

\bibitem{koshelev2}  Koshelev, A.S.:   Stable analytic bounce in non-local Einstein-Gauss-Bonnet cosmology.   	[arXiv:1302.2140 [astro-ph.CO]]

\bibitem{koshelev-v}  Koshelev, A.S.,  Vernov, S.Yu.: Analysis of scalar perturbations in
cosmological models with a non-local scalar field. Class. Quant. Grav. \textbf{28}, 085019 (2011).  [arXiv:1009.0746v2 [hep-th]]

\bibitem{koshelev4}   Koshelev, A.S., Modesto, L.,  Rachwal, L.,  Starobinsky, A.A.:  Occurrence of exact $R^2$ inflation in non-local
UV-complete gravity. [arXiv:1604.03127v1 [hep-th]]

\bibitem{steinhardt}   Lehners, J.-L.,  Steinhardt, P.J.: Planck 2013 results support the cyclic universe.  	arXiv:1304.3122 [astro-ph.CO]

\bibitem{li}  Li, Y-D., Modesto, L.,  Rachwal, L.:  Exact solutions and spacetime singularities in nonlocal gravity. JHEP \textbf{12}, 173 (2015).
[arXiv:1506.08619 [hep-th]]

\bibitem{modesto3}  Modesto, L.: Super-renormalizable quantum gravity. Phys.\ Rev.\ D {\bf 86}, 044005 (2012).
 [arXiv:1107.2403 [hep-th]]

 \bibitem{modesto4}    Modesto, L., Rachwal, L.: Super-renormalizable and finite gravitational theories.
 Nucl. Phys. B {\bf 889}, 228 (2014).  [arXiv:1407.8036 [hep-th]]

\bibitem{modesto}   Modesto, L.,   Tsujikawa, S.:  Non-local massive gravity. Phys. Lett. B \textbf{727}, 48--56 (2013).  	[arXiv:1307.6968 [hep-th]]

\bibitem{moffat}  Moffat, J.M.:  Ultraviolet complete quantum gravity.  Eur. Phys. J. Plus
\textbf{126}, 43 (2011).  	[arXiv:1008.2482 [gr-qc]]

\bibitem{nojiri}  Nojiri, S.,  Odintsov, S.D.: Unified cosmic history in modified gravity: from $F(R)$ theory
to Lorentz non-invariant models. Phys. Rep. \textbf{505}, 59--144 (2011).
 [arXiv:1011.0544v4 [gr-qc]]

\bibitem{nojiri1} Nojiri, S., Odintsov, S.D.: Modified non-local-F(R) gravity as the key for inflation and dark energy. Phys. Lett. B \textbf{659},
821--826 (2008). [arXiv:0708.0924v3 [hep-th]

\bibitem{novello} Novello, M., Bergliaffa, S.E.P.: Bouncing cosmologies.   	Phys. Rep. \textbf{463}, 127--213 (2008).
[arXiv:0802.1634 [astro-ph]]

\bibitem{faraoni} T. P. Sotiriou, V. Faraoni, $f(R)$ theories of gravity. Rev. Mod. Phys.
\textbf{82} (2010) 451--497.  [arXiv:0805.1726v4 [gr-qc]]

\bibitem{stelle}  Stelle, K.S.: Renormalization of higher derivative quantum gravity. Phys. Rev. D \textbf{16}, 953 (1977)

\bibitem{vvz}
Vladimirov, V.S., Volovich, I.V., Zelenov, E.I., p-adic Analysis and Mathematical Physics, 1994

\bibitem{woodard1} Woodard, R.P.: Nonlocal models of cosmic acceleration.      	[arXiv:1401.0254 [astro-ph.CO]]

\bibitem{sasaki}   Zhang, Y.-li.,  Sasaki, M.:  Screening of cosmological constant in non-local cosmology.  Int. J. Mod. Phys. D \textbf{21}, 1250006 (2012).
 [arXiv:1108.2112 [gr-qc]]

\bibitem{zhang}  Zhang, Y.-li.,  Koyama, K.,   Sasaki, M.,   Zhao, G-B.: Acausality in nonlocal gravity theory. JHEP \textbf{1603}, 039(2016).
 [arXiv:1601.03808v2 [hep-th]]
\end{thebibliography}
\end{document}